\documentclass[aps,showpacs,twocolumn]{revtex4}
\usepackage{graphicx}

\newcommand{\BV}{\left(\begin{array}{c}}
\newcommand{\EV}{\end{array}\right)}
\newcommand{\BM}{\left(\begin{array}{cc}}

\newcommand{\RA}{\rightarrow}

\newcommand{\beq}{\begin{equation}}
\newcommand{\eeq}{\end{equation}}
\newcommand{\beqry}{\begin{eqnarray}}
\newcommand{\eeqry}{\end{eqnarray}}
\newcommand{\nonum}{\nonumber}

\begin{document}

%{\huge{}}
%Date 4-05-07

\title{Multichannel oscillations and relations between LSND, \\ KARMEN and MiniBooNE,
with and without CP violation\footnote{This research is supported in part by the Department of Energy under contract
W-7405-ENG-36, the National Science Foundation under NSF Grant \#PHY0099385  and 
the Australian Research Council.}}

\author{T.\ Goldman$^a$}

\affiliation{$^a$Theoretical Division, Los Alamos National Laboratory, Los
Alamos, NM 87545, USA}

\author{G.J.\ Stephenson, Jr.$^b$}

\affiliation{$^b$Dept. of Physics and Astronomy, University of New Mexico, 
Albuquerque, NM 87131, USA}

\author{B.H.J.\ McKellar$^c$}

\affiliation{$^c$School of Physics, University of Melbourne,  Victoria 3010 Australia}

%{\today}\\
\hspace*{0.5in}\\
\begin{flushright}
%\vspace{-1.5in}
\vspace{-0.1in}
{nucl-th/0703023}\\
{LA-UR-07-1479rev}\\
\end{flushright}

\begin{abstract}
We show by examples that multichannel mixing can affect both 
the parameters extracted from neutrino oscillation experiments, 
and that more general conclusions derived by fitting the experimental 
data under the assumption that only two channels are involved in 
the mixing. Implications for MiniBooNE are noted and an example 
based on maximal CP violation displays profound implications for 
the two data sets ($\nu_{\mu}$ and ${\bar{\nu_{\mu}}}$) of that 
experiment. 

\end{abstract}


\vspace{-0.2in}

\maketitle

\section{Introduction}

There has been much discussion\cite{disc,eitel} concerning the difference 
in results between the KARMEN\cite{karmen} and LSND\cite{lsnd} 
experiments regarding appearance of electron antineutrinos, ${\bar 
\nu_{e}}$,  from muon antineutrino, ${\bar \nu_{\mu}}$, sources. Initially, 
the concern was that the mass-squared difference, $\Delta m^{2}$, 
characterizing the oscillation scale does not match up with the differences 
observed in atmospheric\cite{superk} and solar\cite{sno} neutrino 
oscillations (with the latter now both confirmed and superseded by the 
results of the KamLAND\cite{kamland} experiment). This concern was 
based on the assumption that only the three known active neutrino flavors 
participate in the oscillations, so that the third value of $\Delta m^{2}$ is 
determined by the other two values. 

More recently, it has been recognized that light sterile neutrinos may exist 
and participate in oscillation phenomena\cite{us,kusenko,others}. Because 
multiple cycles have not been explicitly observed, this 
raises a serious question regarding an assumption in the analyses of all 
oscillation experiments to date, namely that the oscillation scales are 
sufficiently separated so as not to influence the the values extracted using 
the functional relations in simple, two-channel mixing. We have 
previously shown\cite{us} how a reduced rank see-saw\cite{gmrs, mohap} 
which couples several different oscillations, leads to more complex phenomena, 
as was long ago recognized by Fermi and Ulam\cite{fu} whenever more than 
two oscillators are coupled. 

If there are, in fact, multiple paths which contribute to electron neutrino 
appearance from a muon neutrino source, then the shortest baseline 
(largest $\Delta m^2$) oscillation would appear as excursions from a rising 
longer baseline\cite{us}. The other paths would have independent oscillation 
parameters $\Delta m_{1i}^2$ and $\Delta m_{2j}^2$, where, without 
loss of generality, we treat mass eigenstate "1" is the dominant contributor 
to the initial flavor and "2" as the dominant contributor to the final flavor. Then 
the oscillation from initial to final flavor can be represented as
\beqry
P({\bar \nu_{\mu}}  \RA {\bar \nu_{e}}) & = & A^{2} {\rm sin}^{2}(\Delta m_{12}^{2}x)
 \nonum \\ & + &  B^{2} {\rm sin}^{2}(\Delta m_{1i}^{2}x)
 \nonum \\ & + &  C^{2} {\rm sin}^{2}(\Delta m_{2i}^{2}x)
 + \cdots   \label{fulleqn}
\eeqry
where $x=1.27 \; L/E$ (m/MeV) and the dots indicate that, in principle, more than 
one intermediate channel $i$ may contribute. The coefficients must all be positive 
semi-definite to ensure positivity of the appearance probability. The additional terms  
produce the rising baseline.\cite{us} so that the problem may be viewed as an oscillation 
of the usual two channel type, but occurring over a rising baseline. Of course, for $x$ 
sufficiently small, all of the ${\rm sin}^{2}$ arguments can be simultaneously expanded 
and the appearance probability develops purely quadratically, viz. 
\beqry
P({\bar \nu_{\mu}}  \RA {\bar \nu_{e}}) & \sim & \left[ A^{2}({\Delta m_{12}^{2}})^{2} 
+  B^{2}({\Delta m_{1i}^{2}})^{2}  \right. \nonum \\
& +  & \left. C^{2}({\Delta m_{2i}^{2}})^{2} + \cdots \right] x^{2}  
\eeqry

We provide some simple illustrations, which have the advantage of being 
able to improve the compatibility between the KARMEN and LSND results 
in a way that implicitly makes  predictions for the eagerly awaited results 
from the MiniBooNE\cite{mini} experiment. 

\section{Example with CP Conservation}

We suppose that an oscillation from ${\bar \nu_{\mu}}$ to ${\bar \nu_{e}}$ 
occurs with a value of $\Delta m^{2}$ consistent with the LSND allowed range. 
However, as was shown possible in Ref.\cite{us}, we assume this occurs 
with coupling to other (here unspecified) channels that produces a rising 
baseline for the two channel oscillation. Thus, the probability for detecting 
a ${\bar \nu_{e}}$ of energy $E$ at a distance $L$ from a ${\bar \nu_{\mu}}$ 
source, $P({\bar \nu_{\mu}} \RA {\bar \nu_{e}})$ is given by the explicit part 
of Eq.(\ref{fulleqn}) above, and here we choose some particular examples 
for the parameter values (where we absorb the factor of 1.27 into coefficients 
so that the values in the arguments of the sine functions are $1.27 \times 
\Delta m^{2}$ in units of eV$^2$.):
\beqry
P_{\rm 2ch; High}^{({\bar \nu_{\mu}} \RA {\bar \nu_{e}})} & = & 0.0045 \; {\rm sin}^{2}(0.8 \; L/E) \\
P_{\rm 2ch; Low}^{({\bar \nu_{\mu}} \RA {\bar \nu_{e}})} & = & 0.0600 \; {\rm sin}^{2}(0.2 \; L/E) \\
P_{\rm multich;a}^{({\bar \nu_{\mu}} \RA {\bar \nu_{e}})} & = & 0.005 \; {\rm sin}^{2}(0.7 \; L/E) 
\nonum \\  & +  & 0.001 \; {\rm sin}^{2}(0.3 \; L/E) 
\nonum \\  & +  & 0.0025 {\rm sin}^{2}(0.4 \; L/E) \\
P_{\rm multich;b}^{({\bar \nu_{\mu}} \RA {\bar \nu_{e}})} & = & 0.004 \; {\rm sin}^{2}(0.7 \; L/E) 
\nonum \\  & +  & 0.005 \; {\rm sin}^{2}(0.2 \; L/E) 
\nonum \\  & +  & 0.002 {\rm sin}^{2}(0.5 \; L/E) 
\eeqry
for the appearance rates in the two-channel and multi-channel cases respectively. 
Although the three-channel mass relation is satisfied in this example, we emphasize 
that the intermediate channel need not be an active neutrino (if light sterile neutrinos 
exist\cite{kusenko}) and furthermore, this relation need not have been satisfied if two 
different intermediate channels make the dominant contributions\cite{disc}.  For very 
large $\Delta m^{2}$, rapid oscillations will average to a constant appearance rate 
independent of $L/E$, which we use to set a normalization of $0.0026$ consistent 
with the scale for the signal reported by LSND\cite{lsnd}.   

\begin{figure}
\begin{center}
\includegraphics[width=3in]{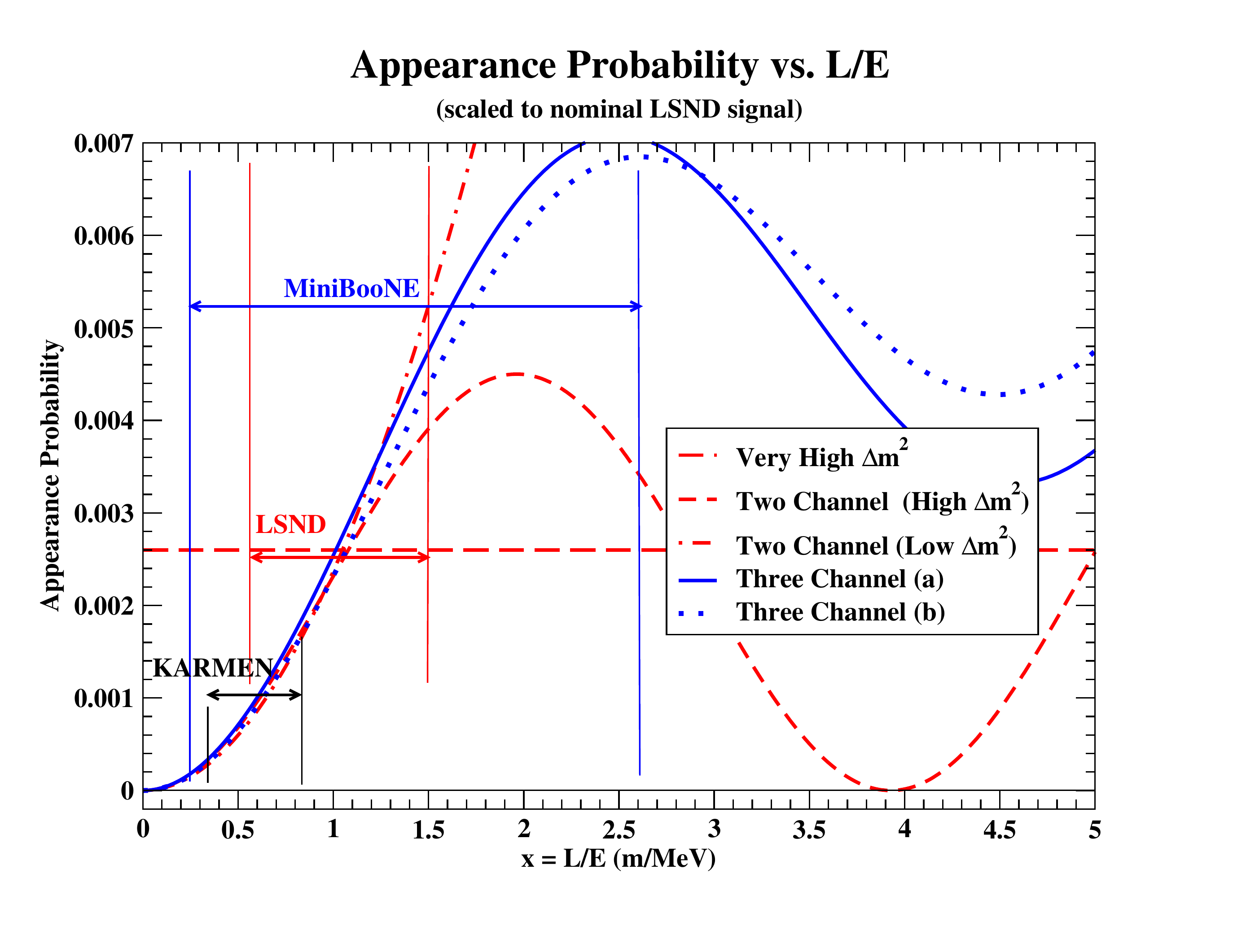}
\end{center}
\caption{\label{LoverE}Two channel mixing ${\bar \nu_{e}}$ appearance 
probabilities from ${\bar \nu_{\mu}}$ for three values of $\Delta m^{2}$ 
compared with the two example three-channel rates discussed in the text 
{\it vs.} $L/E$ .}
\end{figure}

\begin{figure}
\begin{center}
\includegraphics[width=3in]{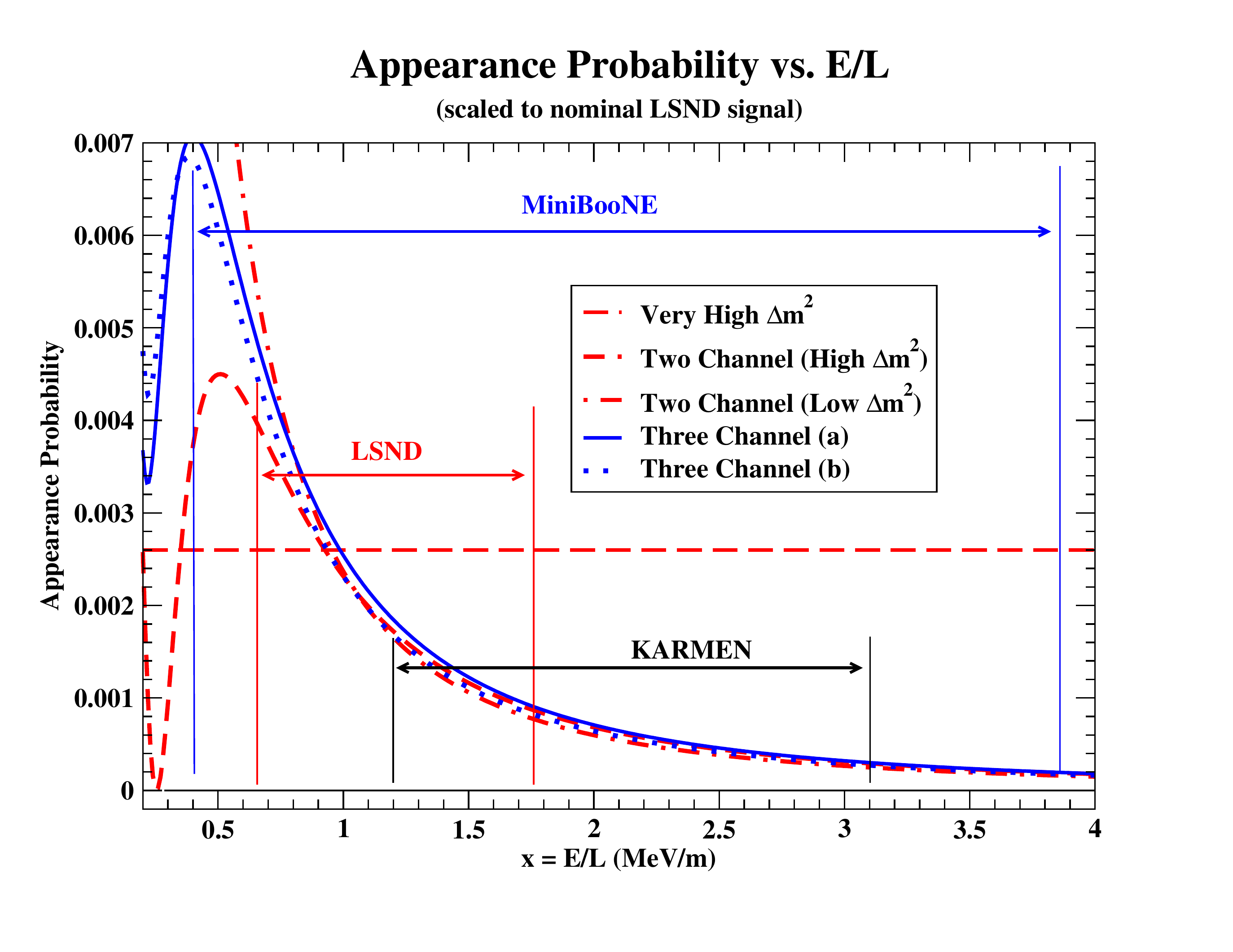}
\end{center}
\caption{\label{EoverL}Two channel mixing ${\bar \nu_{e}}$ appearance 
probabilities from ${\bar \nu_{\mu}}$ for three values of $\Delta m^{2}$ 
compared with the two example three-channel rates discussed in the text 
{\it vs.}  $E/L$ .}
\end{figure}

For the multichannel cases, Fig. \ref{LoverE} reprises the character of the result in 
Fig.(2) of Ref.\cite{us} in the usual $L/E$ terms. Note that the appearance probabilities 
are virtually indistinguishable in the low $L/E$ region covered by the KARMEN and 
LSND experiments, although wide deviations occur in the larger $L/E$ region that 
MiniBooNE can address.  It is also interesting to  consider an  $E/L$ plot of the same 
function as in Eq.(\ref{fulleqn}), along with the corresponding distributions for simple 
two-channel fits to the LSND experiment at high and low values of $\Delta m^{2}$ as 
shown in Fig. \ref{EoverL}. Again, for a very high value of $\Delta m^{2}$, the limited 
resolution results in an averaged, flat distribution, experimentally indistinguishable 
from one independent of $E/L$. 

Both figures include an indication of the range of $L/E$ (or $E/L$) over which 
each of the KARMEN, LSND and MiniBooNE experiments are sensitive. 

%\bigskip

\section{Example with CP Violation}

In the discussion so far, we have not allowed for CP violation. If that occurs,   
additional terms\cite{PDG} arise from the imaginary part of the same product 
of four mixing matrices that produces the positive coefficients above from the 
real part. These terms are of the form
\beqry
\Delta P({\bar \nu_{\mu}}  \RA {\bar \nu_{e}}) & = & D \; {\rm sin}(2\Delta m_{12}^{2}x)
+ E \; {\rm sin}(2\Delta m_{1i}^{2}x) \nonum \\ 
& + &  F \; {\rm sin}(2\Delta m_{2i}^{2}x) + \cdots   \label{CPV}
\eeqry
Note that the constraint of positivity of the coefficients does not apply to these 
terms. In fact, for "maximal" CP violation, in the sense of the conventional angle 
$\delta = \pi/2$, it is straightforward to demonstrate that, depending on which 
way one represents the mass differences, some of the coefficients above must 
be negative semi-definite. Specifically, in the Particle Data Group (PDG) 
formulation\cite{PDG} for exactly three channels
\beqry
D & = &  -E = -F =  s_{12}s_{23}s_{13}c_{23}c_{12}c_{13}^{2}  \nonum \\ 
A^{2} & = &  c_{12}^{2}s_{13}^{2}s_{23}^{2}c_{13}^{2}  \nonum \\  
B^{2} & = &  s_{12}^{2}s_{13}^{2}s_{23}^{2}c_{13}^{2}  \nonum \\  
C^{2} & = &  s_{12}^{2}c_{12}^{2}c_{13}^{2}(c_{23}^{2}-s_{23}^{2}s_{13}^{2})  
 \label{CPcoeffs}
\eeqry
where $s_{12} = {\rm sin}(\theta_{12})$, etc. as usual. Clearly, if any one of 
the conventional CKM/MNS\cite{acronyms} angles vanishes (or reaches 
$\pi/2$) then the CP-violating parts vanish as they must. Note also that the 
positivity of the ${\rm sin}^{2}$ terms is not affected. 

It is the positivity of the appearance probability that requires the relation above 
between $D$, $E$ and $F$ -- these terms all change sign for the CP-conjugate 
channel. Therefore, the sum of the coefficients of $L/E$ for small $L/E$ must 
actually vanish: 
\beq
D \times \left[ \Delta m_{12}^{2} - \Delta m_{1i}^{2} + \Delta m_{2i}^{2} \right] = 0
\eeq
which is guaranteed by the relations among the three mass differences. 

We provide one example with CP violation of an oscillation that agrees with both 
LSND and KARMEN and predicts that the signal in MiniBooNE may be smaller 
than the largest value expected from the LSND results: 
\beqry
P_{\rm CPV}^{({\bar \nu_{\mu}} \RA {\bar \nu_{e}})} & = &
0.0025 \; ({\rm sin}^{2}(1.0 \; L/E))-0.001 \; {\rm sin}(2.0 \; L/E) \nonum \\ 
& + & 0.0005 \; ({\rm sin}^{2}(3.0 \; L/E)) -0.001 \; {\rm sin}(6.0 \; L/E) \nonum \\ 
& + & 0.001 \; ({\rm sin}^{2}(4.0 \; L/E)) + 0.001 \; {\rm sin}(8.0 \; L/E) \nonum \\
& + & 0.01 \; ({\rm sin}^{2}(0.5 \; L/E)) \label{CPVa}
\eeqry
where coefficients of all additional terms are assumed to be negligibly small 
(and we have again absorbed the factor of 1.27 into the numerical parameters). 
 Of course, none of the $\Delta m^{2}$ values 
matches with those inferred from other experiments that do observe flavor 
oscillations, so the scenario here is viable {\em only} in the case that this set 
of oscillations is proceeding through neutrino mass eigenstates that are dominantly 
{\em sterile} neutrino states, with small flavor components. 

For the opposite CP process, applicable to KARMEN and LSND, the 
contributions from the imaginary part of the product of the $U$ matrices 
(the sin($2\Delta m^{2}x$) terms) change sign, so we have
\beqry
P_{\rm CPV}^{(\nu_{\mu} \RA \nu_{e})} & = &
0.0025 \; ({\rm sin}^{2}(1.0 \; L/E)) +0.001 \; {\rm sin}(2.0 \; L/E) \nonum \\ 
& + & 0.0005 \; ({\rm sin}^{2}(3.0 \; L/E)) +0.001 \; {\rm sin}(6.0 \; L/E) \nonum \\ 
& + & 0.001 \; ({\rm sin}^{2}(4.0 \; L/E)) - 0.001 \; {\rm sin}(8.0 \; L/E) \nonum \\
& + & 0.01 \; ({\rm sin}^{2}(0.5 \; L/E)) \label{CPVb}
\eeqry
This formula improves the agreement between KARMEN and LSND,  slightly, 
over the lowest $\Delta m^{2}$ fit to both, and does so without requiring large 
$\nu_{\mu}$ or $\nu_{e}$ disappearance at larger $L/E$ -- the maximum loss 
is only slightly more than 1\%.

\begin{figure}
\begin{center}
\includegraphics[width=3in]{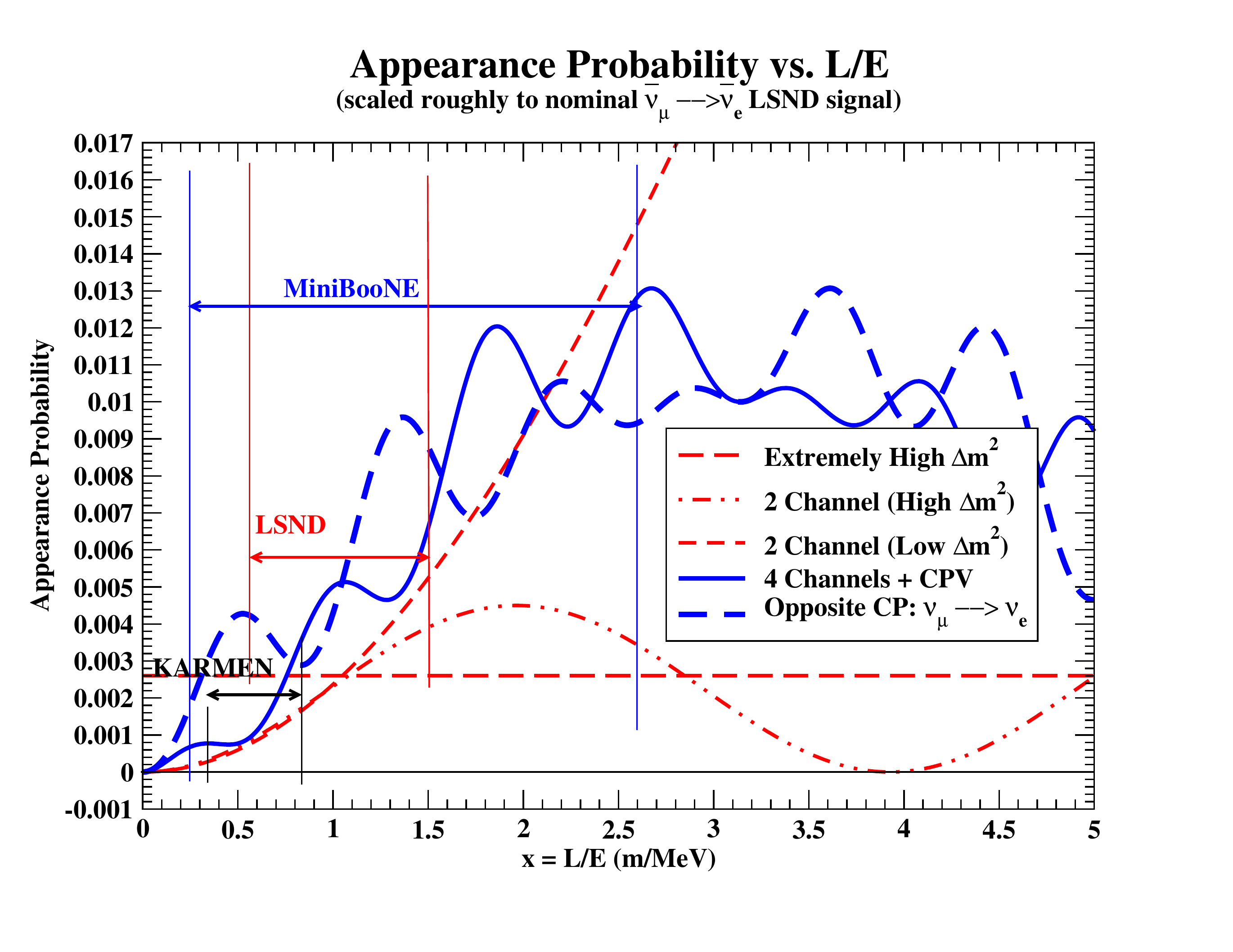}
\end{center}
\caption{\label{CPLoverE}Four channel mixing ${\bar \nu_{e}}$ appearance 
probability from ${\bar \nu_{\mu}}$ for Eq.(\ref{CPVa}) and the appearance 
probability for  the CP conjugate channel for Eq.(\ref{CPVb}) as given in the
 text {\it vs.}  $L/E$ compared with 2 channel descriptions described previously.}
 \label{CPLE}
\end{figure}

\begin{figure}
\begin{center}
\includegraphics[width=3in]{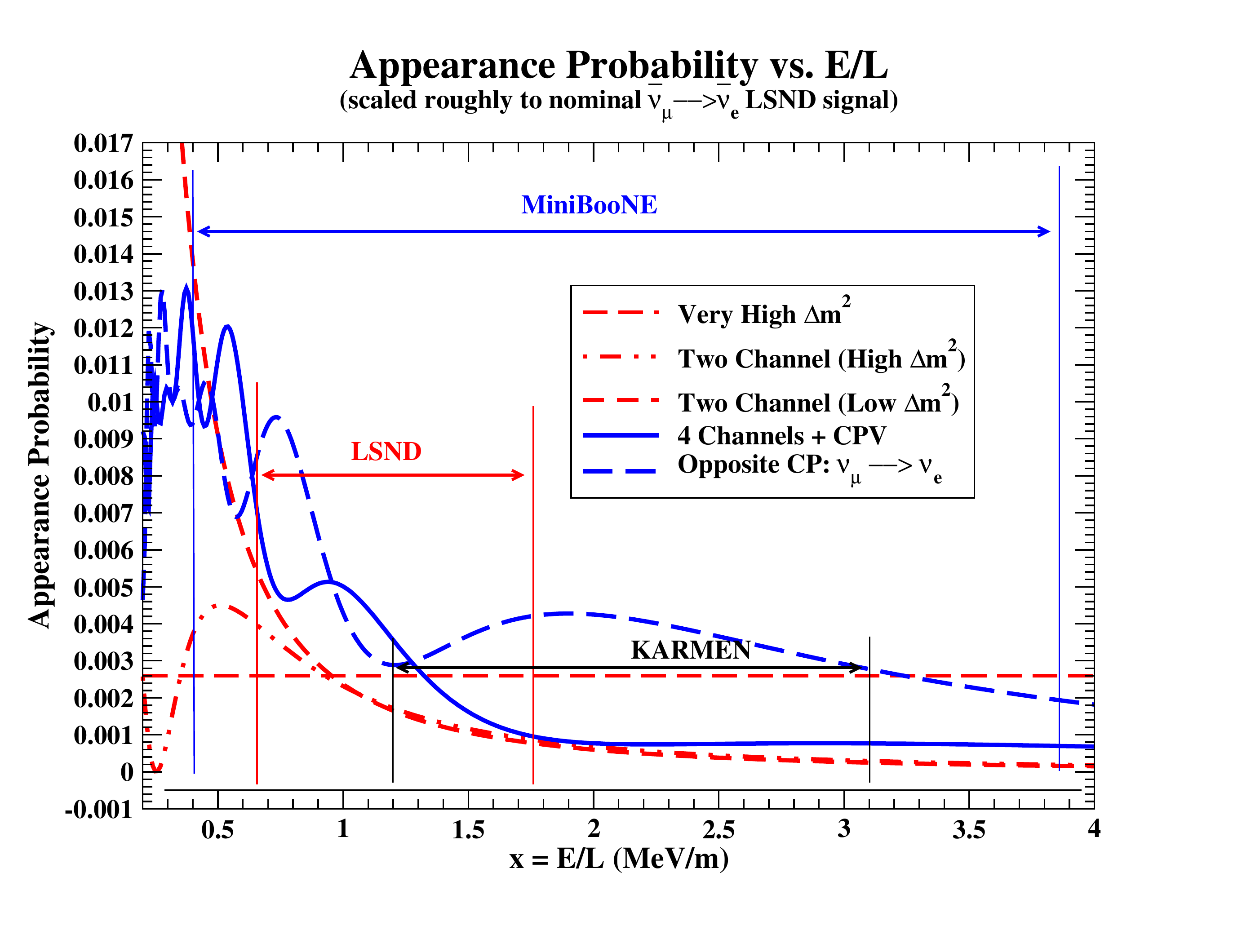}
\end{center}
\caption{\label{CPEoverL}Four channel mixing ${\bar \nu_{e}}$ appearance 
probability from ${\bar \nu_{\mu}}$ for Eq.(\ref{CPVa}) and the appearance 
probability for  the CP conjugate channel for Eq.(\ref{CPVb}) as given in the
 text {\it vs.}  $E/L$ compared with 2 channel descriptions described previously.}
 \label{CPEL}
\end{figure}

We plot the formulae in Eqs.(\ref{CPVa}) and (\ref{CPVb}) vs. $L/E$ and 
$E/L$ in Figs. \ref{CPLE} and \ref{CPEL} for comparison with the CP-conserving 
oscillations shown earlier. As is apparent both from these Figures and 
from the crude $\Delta m^{2}$ values, these formulae do not represent 
a "fit" to the data, but are simply an indication of the possibilities available 
if one does not arbitrarily constrain the entire neutrino oscillation picture 
to three active Majorana neutrino flavor and (light) mass eigenstates. 

\section{Discussion}

The form in Eq.(\ref{fulleqn}) and the additional terms in Eq.(\ref{CPV}) have 
far too many parameters to be tightly fit with data available from present 
neutrino experiments, hence the predeliction for fitting to two-channel mixing 
scenarios. (Even when more channels are attempted, the dominance of one 
scale has been assumed\cite{superk} or the two-channel fit results are 
used\cite{disc}.) However, since the rising baseline we observed possible\cite{us} 
is roughly quadratic, corresponding to the opening of contributions from a longer 
wavelength oscillation, it should be viable to include the possibility of a rising 
baseline with one additional parameter, $T$, viz., 
\beqry
P({\bar \nu_{init}} \RA {\bar \nu_{final}}) & \sim  & A^{2}{\rm sin}^{2}(1.27\Delta m^{2}L/E)
\nonum \\ & & + T(1.27\Delta m^{2}L/E)^{2}  + \cdots 
\eeqry
without CP violation, where $T$ accounts for the rising baseline/additional 
channels, or even more compactly, 
\beqry
P({\bar \nu_{init}} \RA {\bar \nu_{fin}}) & \sim  & S(1.27\Delta m^{2}L/E)^{2}
\nonum \\ & & + V(1.27\Delta m^{2}L/E)^{3} + \cdots \label{tayloreqn}
\eeqry
where $S$ absorbs all of relative amplitudes and ratios of $\Delta m^{2}$ 
values of longer wavelength channels into one parameter and $V$ absorbs 
all of the additional parameters for any number of CP-violating effects into 
another. In fact, these functional forms do not even depend on our initial 
physical ansatz , Eq.(\ref{fulleqn}), and have only the disadvantage of not 
being applicable for large values of $L/E$.  We recommend that all oscillation 
experiments test such functional forms to determine whether or not the 
$\chi$-squared per degree of freedom of the fit to their data is, or is not, 
improved by such additions to the standard two-channel analysis. 

As noted in Ref.\cite{eitel}, with two-channel mixing, the LSND and KARMEN 
experiments are in best agreement for low values of $\Delta m^{2}$ because 
in that case, the difference in distances affects the results quadratically in favor 
of LSND, whereas for very high values of $\Delta m^{2}$ they should have seen 
the same size signal and hence are in disagreement. However, here we see that, 
while even for an intermediate value of $\Delta m^{2}$ the agreement would 
be marginal in two-channel mixing, the problem is reduced once the effect of 
a third channel on the baseline for the two-channel oscillation is included. The 
improvement is even more striking when CP violation is allowed. In fact, the ratio 
between the signal expected in the two experiments can achieve essentially the 
same value (or an even better one) as that obtained with a small $\Delta 
m^{2}$ fit. 

The concern over a smaller value of $\Delta m^{2}$ for LSND is the effect of the 
larger intrinsic mixing amplitude required to match the data obtained in the region 
of small $L/E$: It predicts large effects, particularly disappearance rates, at much 
larger values of $L/E$ typical of reactor experiments\cite{CHOOZ}, for instance. 
However, as our  examples demonstrate, the rising baseline breaks the relation, 
seen in two-channel mixing, between the rate that an appearance signal increases 
with increasing $L/E$ and the size of the signal in the initial range of the effect. As 
shown in our examples, the total appearance rate remains near 1\% at all values 
of $L/E$, completely consistent with the limits from short baseline disappearance 
experiments\cite{CHOOZ,CDHS} for both $\nu_{\mu}$ and $\nu_{e}$. We emphasize 
that this is true even though a two-channel fit would require a much larger overall 
amplitude in order for the signal to have grown to the size reported by LSND  yet 
have remained too small to be observed by KARMEN with its shorter baseline (which 
one would think not likely to be significantly shorter). 

Finally, we note that the inclusion of CP violation, which is to be expected, (but 
not CPT violation, which would be revolutionary) further improves agreement 
between KARMEN and LSND and makes explicit testable predictions for the 
results of MiniBooNE, as long as the possibility of light sterile neutrinos is 
allowed. We reiterate that small amplitude mixing to light sterile neutrinos poses 
no conflict with any known laboratory experimental data. 

We conclude that only when oscillation experiments can all provide unbiased 
$L/E$ distributions, rather than reporting parameters for two-channel fits to the 
oscillations observed, will definitive conclusions be possible, regarding neutrino 
mass and mixing parameters, that are independent of theoretical biases. 

The examples we have presented also suggest that, at low energy, the MiniBooNE 
experiment may observe a considerably larger {\em or} smaller signal for $\nu_{e}$ 
appearance than would be expected from the two-channel fits to KARMEN and LSND. 
However,  the examples also show that this would {\em not} necessarily contradict the 
results of either of those two experiments, whether considered separately or jointly. 

We thank Bill Louis and Geoff Mills for several discussions on this subject.

\end{document}